\documentclass[9pt,twocolumn,twoside]{opticajnl}
\journal{optica}
\setboolean{shortarticle}{true}
\usepackage{siunitx}

\title{UNWRAP: Single-shot phase analysis of bright squeezed vacuum by Newton-ring interferometry}

\author[1,2,*]{Haoran Ma}
\author[3,4]{Yihe Wang}
\author[1,2,5]{Chuan Cheng}
\author[2,6,7]{Ian Gabalski}
\author[2,6]{Aaron M.~Ghrist}
\author[1,2]{Eleanor Weckwerth}
\author[4,8]{Mohammed Th.~Hassan}
\author[1,2,6]{Philip H. Bucksbaum}

\affil[1]{Department of Physics, Stanford University, Stanford, California 94305, USA}
\affil[2]{Stanford PULSE Institute, SLAC National Accelerator Laboratory, Menlo Park, California 94025, USA}
\affil[3]{Department of Electrical and Computer Engineering, University of Arizona, Tucson, Arizona 85721, USA}
\affil[4]{Department of Physics, University of Arizona, Tucson, Arizona 85721, USA}
\affil[5]{State Key Laboratory for Mesoscopic Physics and Frontiers Science Center for Nano-Optoelectronics, School of Physics, Peking University, Beijing 100871, China}
\affil[6]{Department of Applied Physics, Stanford University, Stanford, California 94305, USA}
\affil[7]{Department of Chemistry, University of California, Davis, Davis, California 95616, USA}
\affil[8]{James C. Wyant College of Optical Sciences, University of Arizona, Tucson, Arizona 85721, USA}

\affil[*]{mahr@stanford.edu}

\begin{abstract}
The mean field of pulsed bright squeezed vacuum (BSV) vanishes, but the variance can be large, and is strongly phase-dependent.
We employ ultrafast Newton-ring wavefront retrieval and analysis of phase (UNWRAP) to extract the field amplitude and phase profile of each BSV pulse, 
by Fourier analysis of the interference rings between the BSV and a coherent reference with different phase-front curvature. 
The concentric ring interference fringes are robust against pointing mismatch and drift, thereby simplifying the analysis. 
The measured distribution has two phase peaks separated by $\pi$ radians. 
The rings sweep the delay between the two pulses, so a single frame can capture the temporal structure of the BSV field, even when it is composed of several temporal modes.
\end{abstract}

\setboolean{displaycopyright}{false}

\begin{document}
\maketitle

Bright squeezed vacuum (BSV), generated by high-gain parametric down-conversion, combines macroscopic pulse energies with large photon-number fluctuations~\cite{Iskhakov2012,Chekhova2015}. It differs from a coherent field not only in the magnitude of its noise but in its phase dependence: the noise is reduced in one quadrature and amplified in the other. Recent experiments have shown that the strong photon-number noise of BSV enhances multiphoton processes~\cite{Spasibko2017} and modifies strong-field processes such as high-harmonic generation~\cite{Rasputnyi2024}, electron emission from metal needle tips~\cite{Heimerl2024} and multiphoton atomic ionization~\cite{Jiang2026,LiuXe2026}. These measurements, along with theory~\cite{Gorlach2023,RiveraDean2025,Schimmoller2026}, establish that strong-field processes are sensitive to the photon statistics of the driving light, but they do not show how the phase-dependent noise that distinguishes BSV from thermal light of the same brightness affects them. Resolving it requires the field of each pulse relative to a phase reference.

Homodyne detection provides such a reference: it measures the quadratures of a field through its interference with a coherent local oscillator~\cite{Lvovsky2009}. For ultrafast BSV, spectral interferometry with a fully characterized reference pulse has retrieved the pulse shape of each shot and identified its $\pi$ phase ambiguity~\cite{Kern2026}, and sub-cycle field sampling with a nonlinear gate has resolved the randomly signed bursts within a pulse~\cite{GholamMirzaei2026}. Here we encode the field of each BSV pulse in the Newton rings it forms with a coherent reference, and read out its amplitude, phase and temporal profile from one camera frame at the 1-kHz repetition rate of the laser.

\begin{figure*}[t]
\centering
\includegraphics[width=\textwidth]{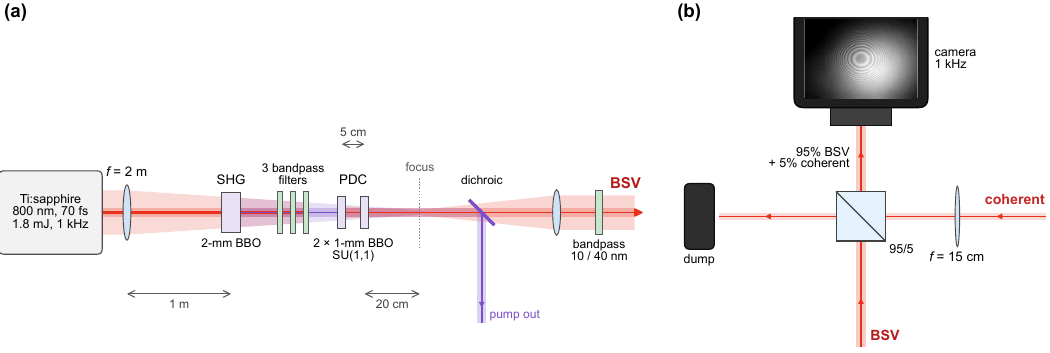}
\caption{Experimental layout. (a) Generation of the BSV. The 800-nm pulses are focused by a 2-m focal-length lens. A 2-mm BBO crystal 1~m downstream generates the second harmonic (SHG), which pumps parametric down-conversion (PDC) in two 1-mm BBO crystals 5~cm apart, about 20~cm before the focus. A dichroic mirror and three bandpass filters between the SHG and PDC crystals ensure that no residual 800-nm light enters the PDC crystals. A dichroic mirror behind the PDC crystals removes the pump, a second lens recollimates the BSV, and a bandpass filter, 10 or 40~nm wide, sets its bandwidth. (b) Newton-ring readout. The BSV and the coherent reference, focused by a 15-cm focal-length lens, are combined on a 95/5 beam splitter, and a CMOS camera (MindVision MV-XGC51GC/M) records the interference rings. The screen shows a measured frame.}
\label{fig:setup}
\end{figure*}

\begin{figure}[t]
\centering
\includegraphics[width=\linewidth]{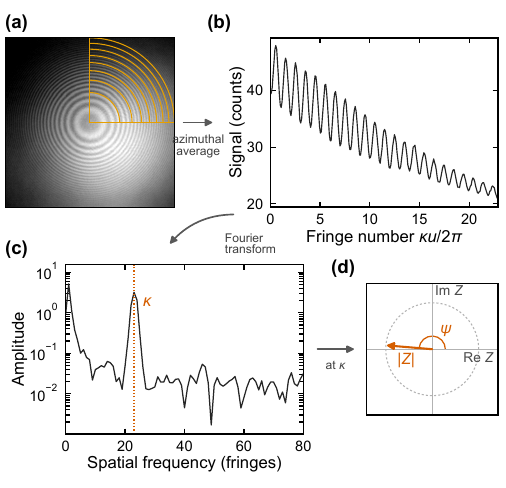}
\caption{Readout of one frame. (a) One camera frame of the interference between the BSV and the reference; the arcs mark every 32nd of the radial bins of equal width in the squared radius $u=r^2$. (b) Azimuthal average of the frame against the fringe number $\kappa u/2\pi$. (c) Spatial spectrum of the profile on a logarithmic scale; $\kappa$ marks the carrier. (d) The complex Fourier coefficient at $\kappa$, $Z=|Z|e^{i\psi}$.}
\label{fig:method}
\end{figure}

\textit{Experimental setup.} The output of a 1-kHz Ti:sapphire amplifier (800~nm, 70~fs, 1.8~mJ) is split into two beams. One beam generates the BSV by second-harmonic generation (SHG) in a BBO crystal and parametric down-conversion (PDC) in two BBO crystals [Fig.~\ref{fig:setup}(a)]; the other beam is the coherent reference. The amplifier pulses are left longer than their transform limit, so that their chirp pre-compensates the dispersion of the crystals. The two PDC crystals form an SU(1,1) interferometer~\cite{Lemieux2016}. The reference meets the BSV on the camera with a different wavefront curvature [Fig.~\ref{fig:setup}(b)], and their interference produces Newton rings [Fig.~\ref{fig:method}(a)]. Each frame is averaged over azimuth in bins of equal width in the squared radius $u=r^2$, in which the rings have a single spatial frequency $\kappa$ [Fig.~\ref{fig:method}(b)]. The Fourier transform of this profile~\cite{Takeda1982,Ge2001} [Fig.~\ref{fig:method}(c)] separates the carrier from the static background and the beam envelopes, which occupy only the lowest spatial frequencies, and its complex coefficient at $\kappa$, $Z_j=|Z_j|e^{i\psi_j}$, is the fringe amplitude and phase of frame $j$ [Fig.~\ref{fig:method}(d)]. With the intensity of the reference approximately constant from shot to shot, $|Z_j|$ is proportional to the amplitude of the BSV field of pulse $j$ and $\psi_j$ is its phase relative to the reference.

The phase $\psi_j$ is insensitive to the pointing jitter of the two beams. A tilt $\boldsymbol\alpha$ between the two wavefronts adds a linear phase $k\boldsymbol\alpha\cdot\mathbf r$ to the quadratic phase $\kappa r^2$ of the rings. Completing the square, the rings keep their spacing $\kappa$ and their centre moves by $\mathbf q=k\boldsymbol\alpha/2\kappa$, 21 pixels per mrad in the frame of Fig.~\ref{fig:method}. A displacement of either beam moves the ring centre in the same way. Pointing therefore shifts the rings as a whole, by $\mathbf q$ from the calibrated centre of the azimuthal average. Over the azimuth the linear phase averages to the real factor $J_0(2\kappa rq)$. A tilt lowers the fringe contrast, by 12\% for $q$ of 1 pixel (a tilt of 48~\textmu rad), but does not change the extracted phase at first order; the second-order change is at most of order $\kappa q^2$, $0.1^\circ$ at 1 pixel. In a crossing-angle geometry, by contrast, the same tilt changes the fringe spacing itself. The sub-cycle field sampling of Ref.~\cite{GholamMirzaei2026} uses this geometry and tracks the path-length jitter between shots with a separate pair of reference beams on the same sensor. In the Newton-ring geometry a tilt leaves the ring spacing unchanged, and the phase of each shot, path jitter included, is read from the same fringes that carry the field; the jitter is separated from the state statistically, as described below.

\begin{figure}[t]
\centering
\includegraphics[width=\linewidth]{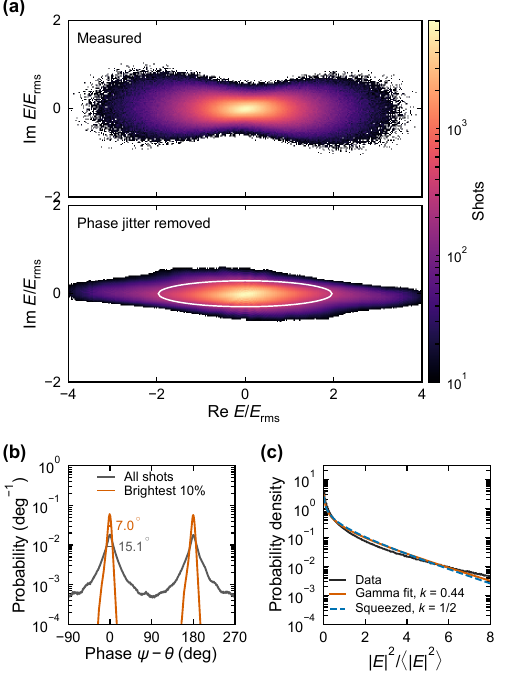}
\caption{Field statistics of the BSV sampled by the camera, behind the 10-nm bandpass filter. (a) Husimi function of the BSV field: real and imaginary parts of the field relative to the squeezing axis, in units of its rms magnitude; shots per bin on a logarithmic colour scale. Top, as measured, no shot discarded. Bottom, after ring-by-ring angular deconvolution of the $10.1^\circ$ rms phase jitter between the two beams. The ellipse has semi-axes $2a$ and $2b$, twice the rms widths $a=0.98$ and $b=0.147$ along the noisy and quiet axes. (b) Distributions of the phase relative to the squeezing axis, $\psi-\theta$, after the same jitter removal, for all shots and for the brightest 10\% by fringe amplitude; the bars mark the FWHM of the peaks, $15.1^\circ$ and $7.0^\circ$. (c) Distribution of the fringe intensity $|E|^2$ normalized to the local mean intensity, with the gamma fit ($k=0.44$) and the gamma distribution of a single squeezed mode ($k=1/2$)~\cite{Iskhakov2012,Perez2014}.}
\label{fig:state}
\end{figure}

\textit{Field statistics.} The squeezing axis $\theta=\tfrac12\arg\langle e^{2i\psi}\rangle$ is evaluated in blocks of 7~s, which follows the slow drift of the relative phase, and $E=Ze^{-i\theta}$. Figure~\ref{fig:state}(a), top, shows the measured distribution of the field $E$ relative to the squeezing axis, for the BSV behind the 10-nm filter. Following Ref.~\cite{Rasputnyi2026}, we refer to this distribution as the Husimi function of the BSV. The distribution is centred on the origin and elongated along $\operatorname{Re}E$, with $\mathrm{rms}(\operatorname{Re}E)/\mathrm{rms}(\operatorname{Im}E)\simeq4.4$. Its shape, however, is a bowtie rather than the ellipse of a squeezed vacuum: the transverse width grows with $|\operatorname{Re}E|$, as for a squeezed state whose phase has diffused~\cite{Franzen2006}. We attribute the diffusion to the technical phase jitter between the two beams, about $10^\circ$ rms, which rotates each shot about the origin. We measure the jitter from the angular width of the distribution at large $|E|$, where the transverse width of the state contributes least, and remove it by deconvolution in the angle, ring by ring. With the jitter taken as independent of the amplitude, the deconvolved distribution [Fig.~\ref{fig:state}(a), bottom] is an ellipse with rms widths of $0.98$ along and $0.147$ across the axis, a ratio of about $6.7$.

The phase distribution [Fig.~\ref{fig:state}(b)] has two sharp peaks. For the brightest 10\% of shots they are exactly $\pi$ apart ($179.9^\circ\pm0.1^\circ$), as expected for the two signs of the amplified quadrature of BSV~\cite{Kern2026}. The peaks are almost equally populated, $50.3{:}49.7$, and have a FWHM of $7.0^\circ$ once the jitter is removed. Between the peaks the probability falls by more than three orders of magnitude, so every bright shot can be assigned to one of the two signs of the field.

The fringe intensity $|E|^2$, normalized to the slowly drifting mean, follows a gamma-shaped distribution [Fig.~\ref{fig:state}(c)]. A gamma fit to the bulk of the distribution gives a shape parameter $k\simeq0.44$, close to the $k=1/2$ of a single squeezed mode~\cite{Iskhakov2012,Perez2014,Chekhova2015}; the tail is heavier than the fit, as expected if the parametric gain also fluctuates from shot to shot.

\begin{figure*}[t]
\centering
\includegraphics[width=\textwidth]{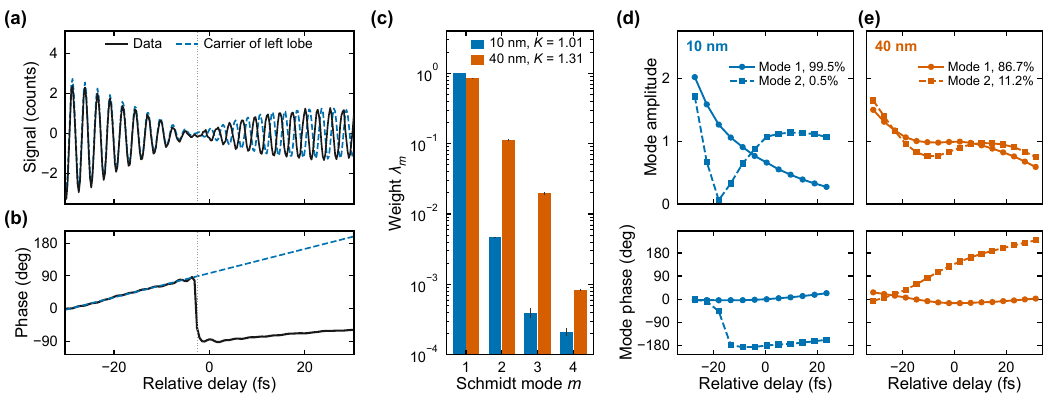}
\caption{Temporal modes of the BSV. (a) Fringe profile of a single pulse behind the 40-nm filter, minus its local mean, against the relative delay of the two pulses, with the carrier of the left lobe continued across the node. (b) Local fringe phase of the same pulse, demodulated over 2.35 fringes and relative to the ensemble phase profile; dashed, the linear fit over the left lobe that sets the carrier in (a). (c) Schmidt weights $\lambda_m$ of the cross-correlation covariance behind the 10-nm filter and the 40-nm filter; $K=1/\sum_m\lambda_m^2$; error bars from a block bootstrap. (d), (e) Amplitude and phase of the first two Schmidt modes against the relative delay, behind the 10-nm and the 40-nm filter. Amplitudes are normalized to unit rms over the slices; the phase of mode 2 is referred to mode 1.}
\label{fig:multimode}
\end{figure*}

\textit{Temporal modes.} The relative phase advances by one fringe per optical cycle, 2.67~fs at 800~nm, so the 23 fringes across the aperture also sweep the relative delay of the two pulses, by about 60~fs. A single frame is therefore a spatially encoded cross-correlation of the BSV with the reference, as in the sub-cycle field sampling of Ref.~\cite{GholamMirzaei2026}, with the delay encoded in the wavefront curvature instead of a crossing angle. With a characterized reference pulse the field across the pulse could be retrieved from it by deconvolution, as in Ref.~\cite{Kern2026}, and attosecond metrology of the BSV field has been proposed~\cite{Stammer2026}; here the cross-correlation itself is used, which already carries the temporal information of the pulse. Demodulating the profile over a window of about two fringes gives its local amplitude and phase along the delay axis. Figure~\ref{fig:multimode}(a) shows the fringe profile of one pulse behind the 40-nm filter against the relative delay, and Fig.~\ref{fig:multimode}(b) its local phase. The fringes vanish at a relative delay of $-3$~fs, to 2\% of their amplitude in the neighbouring lobe, and recover on either side into two lobes about 55~fs apart. Across the node the phase jumps by $165^\circ$, so the right lobe is out of phase with the carrier that continues the left lobe. If the BSV occupied a single temporal mode, every pulse would have the same profile, scaled by its own complex field: the ensemble envelope, which is the rms envelope over all pulses, and the ensemble phase up to a constant. The ensemble envelope has no node, so the node and the sign change belong to this pulse, whose field is a superposition of at least two temporal modes of comparable amplitude, as in the bursts of Ref.~\cite{GholamMirzaei2026} and the nodes of Ref.~\cite{Kern2026}.

To resolve these modes, we decompose the cross-correlation into its Schmidt modes, the eigenfunctions of its first-order correlation function~\cite{Christ2011}, as has been done for the spatial and spectral modes of BSV from the covariance of single-shot intensity distributions~\cite{Barakat2025,LiuBSV2026,LiTomo2026}. Each profile is cut into non-overlapping delay slices of about 1.7 fringes, each slice $s$ is demodulated at the carrier to a complex amplitude $z_s$, and the covariance $\langle z_s z_{s'}^*\rangle$ over the shots is diagonalized. Its eigenvectors are the Schmidt modes of the BSV as seen through the reference pulse, and the normalized eigenvalues $\lambda_m$ are their weights [Fig.~\ref{fig:multimode}(c)]; the first two modes are shown in Fig.~\ref{fig:multimode}(d),(e).

Behind the 10-nm filter the first mode holds 99.5\% of the weight and the second 0.5\%, so the Schmidt number $K=1/\sum_m\lambda_m^2$ is 1.01. Behind the 40-nm filter the BSV is multimode, as expected for broadband down-conversion from two crystals~\cite{Lemieux2016}: the first three modes have weights of 86.7\%, 11.2\% and 2.0\%, and $K=1.31$. At both bandwidths the first mode follows the ensemble envelope and has a smooth phase [Fig.~\ref{fig:multimode}(d), (e)]. Its phase varies by less than $30^\circ$ (10~nm) and $45^\circ$ (40~nm) across the window and by at most $7^\circ$ between neighbouring slices, so the dominant mode carries little chirp relative to the reference and no phase jump. The structure sits in the higher modes. Behind the 10-nm filter the second mode has a node and a $\pi$ phase step. Behind the 40-nm filter it has no node inside the window, and its phase advances by about $230^\circ$ relative to the first mode across the 60~fs: it is displaced from the first mode mainly in carrier wavelength, by about 20~nm, so it passes the 40-nm filter but not the 10-nm filter. The reference pulse filters the cross-correlation to its own bandwidth, so these values of $K$ count only the modes that the reference resolves within the window.

Because the readout is single-shot, multimode pulses such as that of Fig.~\ref{fig:multimode}(a) can be identified and removed after the fact, and an experiment driven by the BSV can be post-selected on shots with a single dominant mode. In addition, a larger curvature mismatch, on a sensor with more pixels, would extend the sweep to about 200~fs and cover the whole pulse in every frame.

In summary, we have developed UNWRAP, a Newton-ring interferometer that reads out the amplitude, phase and temporal profile of every BSV pulse from a single camera frame, at the repetition rate of the laser and robustly against pointing jitter and phase drift. With it we have characterized the Husimi function of the BSV and its temporal modes. Because the method is simple, it can run on a pick-off of the beams beside a strong-field experiment, characterize every BSV pulse that drives the main experiment, and so provide the reference against which the shots of the main experiment are sorted~\cite{Ma2026photoelectrons}.

\begin{backmatter}
\bmsection{Funding} UK Research and Innovation--National Science Foundation joint program ``Exploiting Quantum Information Science in Chemistry''; National Science Foundation (PHY-2309238); U.S. Department of Energy, Office of Science, Basic Energy Sciences, Chemical Sciences, Geosciences, and Biosciences Division.

\bmsection{Acknowledgment} H.M., Y.W., M.H., and P.B. were supported by the UK Research and Innovation--National Science Foundation joint program, C.C., A.G., and E.W. by the National Science Foundation, and I.G. by the Department of Energy BES CSGB AMOS program. We acknowledge critical discussions with the APIQuL collaboration, especially Carla F.d.M. Faria, Philipp Stammer, and Javier Rivera-Dean.

\bmsection{Disclosures} The authors declare no conflicts of interest.

\bmsection{Data availability} Data underlying the results presented in this paper
may be obtained from the authors upon reasonable request.

\end{backmatter}

\bibliography{refs}
\bibliographyfullrefs{refs}
\end{document}